\documentclass[useAMS,usenatbib]{mn2e}

\usepackage{times}
\usepackage{txfonts}
\usepackage{bm}

\usepackage{color}
\usepackage{graphicx}
\usepackage{hyperref}

\newcommand{\apj}{ApJ}
\newcommand{\aj}{AJ}

\newcommand{\Ds}{D_{\rm s}}
\newcommand{\zs}{z_{\rm s}}

\newcommand{\Dds}{D_{\rm ds}}
\newcommand{\zd}{z_{\rm d}}

\newcommand{\thetaE}{\theta_{\rm E}}
\newcommand{\thetas}{\theta_{\rm s}}

\begin{document}

\title{Moderate Galaxy-Galaxy Lensing}

\author[Mao, Wang and Smith]{Shude Mao$^{1,2}$\thanks{shude.mao@gmail.com; jian-w06@mails.tsinghua.edu.cn}, Jian Wang$^{3}$, Martin C. Smith$^{1,4}$
 \\
$^1$ National Astronomical Observatories, Chinese Academy of Sciences, 20A Datun Road, Beijing 100012, China \\
$^2$ Jodrell Bank Centre for Astrophysics, the University of Manchester, Manchester, M13 9PL, UK \\
$^3$ Department of Physics and Tsinghua Center for Astrophysics, Tsinghua University, Beijing 100084, China\\
$^4$ The Kavli Institute for Astronomy and Astrophysics, Peking University, Yi He Yuan Lu 5, 
Beijing 100871, China
}
\date{Accepted ........Received .......; in original form ......}
\pubyear{2011}
\maketitle
\begin{abstract}
We study moderate gravitational lensing where a background galaxy is
magnified substantially, but not multiply imaged, by an intervening
galaxy. We focus on the case where both the lens and source are elliptical galaxies.
The signatures of moderate lensing include isophotal distortions and 
systematic shifts in the fundamental plane and Kormendy relation, which can potentially be used to statistically determine the galaxy mass profiles. These effects are illustrated using Monte Carlo simulations of galaxy pairs where the foreground galaxy is modelled as a  singular isothermal sphere model and observational parameters appropriate for the Large Synoptic Survey Telescope (LSST). 
The range in radius probed by moderate lensing will be larger than that by strong lensing, and is in the interesting regime where the density slope may be changing. 
\end{abstract}


\section{Introduction}
\label{sec:introduction}

Gravitational lensing is usually divided into microlensing, strong lensing,
and weak gravitational lensing. Microlensing refers to
the temporal change of magnification of a background source lensed
by an intervening object (usually stars). Strong lensing occurs when
multiple images or strong distortions (e.g. giant arcs) of a
background source are caused by an intervening galaxy or cluster.
For weak lensing the distortion of a background source by the
lensing object is much more subtle. For example, a circular source will be lensed
into an ellipse with ellipticity of a few percent. Such subtle
deviations have to be inferred statistically 
by averaging over a large number of background galaxies.
All these fields in gravitational lensing have been extensively
studied, with diverse applications ranging from cosmology to the detection of extrasolar planets (for a review on these topics, see \citealt{skw06} and references therein). 

In this work, we shall explore the
intermediate regime which we term as ``moderate gravitational lensing''. In
this case, the magnification is still significant, but no
multiple images occur.  In the context of clusters of galaxies, \cite{wil98} and \cite{fut98} have considered a class of 
gravitationally lensed, highly magnified, yet morphologically regular images
(originally motivated by observations of the object cB58, \citealt{yee96, sei98}).
More recently, \citet{son11} discussed the magnification effect in
with a focus on how to remove the mass-sheet degeneracy. In this paper, we shall explore the case where an elliptical galaxy is lensed by a foreground elliptical galaxy.
 Many cases are expected to be found in future surveys
(e.g. by PANSTARRS\footnote{http://pan-starrs.ifa.hawaii.edu/public/home.html} and
LSST\footnote{http://www.lsst.org}) where hundreds of millions of galaxies will
be imaged. Many pairs of galaxies that are close to each other but at
different redshifts will be discovered. Most of these will not be
multiply-imaged. At very large separations, a weak-lensing
galaxy-galaxy analysis will be appropriate; at very small separations, multiple images may form.  In this paper we carry
out Monte Carlo simulations of moderate galaxy-galaxy lensing at intermediate separations to illustrate the signatures, and what we can learn from these. 

The paper is organised as follows. In Section
\ref{sec:model}, we describe the simple singular isothermal sphere model we use. In Section
\ref{sec:result}, we present the main results, including the
predictions for the optical depth, isophotal distortions, and systematic offsets in the fundamental plane and Kormendy relation. We present a summary and discussion in
Section \ref{sec:conclusion}.  Throughout this paper, we adopt a flat $\Lambda$CDM cosmology
model with  $\Omega_{\rm m, 0} =0.27$ and $\Omega_{\Lambda,0} = 0.73$ for the matter and cosmological constant, and the Hubble constant is written as $H_0 = 100h~ \rm km~s^{-1}~Mpc^{-1}$ with $h=0.705$ (\citealt{kom09}).

\section{The Lens and Source Model}
\label{sec:model}

In this section, we describe the lens mass model, the source
population and the surface brightness models of foreground and background galaxies
which we will use for Monte Carlo simulations of galaxy pair catalogues. 

\subsection {Lens mass model}

For simplicity, we model the mass profile of the lensing galaxy as a singular
isothermal sphere (SIS). This simple model is analytically tractable, and
fits the observational data well (e.g., \citealt{koo09}). The surface
mass density can be modelled as  
\begin{equation}
\Sigma(\xi) = {\sigma^2 \over 2G} {1 \over \xi}, \label{eq:dis}
\end{equation}
where $\xi$ is the (physical) distance in the lens plane, and
$\sigma$ is the one-dimensional velocity dispersion. The angular
Einstein radius of an SIS is given by
\begin{equation} \label{eq:ein}
\thetaE = 4 \pi {\sigma^2 \over c^2} {\Dds \over \Ds}=1.^{''}4
\left({\sigma \over 220~\rm km~s^{-1}}\right)^2  {\Dds \over \Ds},
\end{equation}
where $c$ is the speed of light, $\Dds$ is the angular diameter
distance between the lens and the source, and $\Ds$ is the angular
diameter distance to the source.

A point source at unlensed angular position $0< \thetas < \thetaE$ from the galaxy
produces two images at the angular positions $\theta_{1, 2} = \thetas \pm 
\thetaE$. The magnification of each image and the total absolute magnification are given by
\begin{equation} \label{eq:magmu}
\mu_{1, 2} = \frac{\theta_{1, 2}}{\thetas}, ~~~ \mu_{\rm total} = |\mu_1|+|\mu_2|=2\frac{\thetaE}{\thetas}, ~~
0<\thetas<\thetaE.
\end{equation}
Notice that the total magnification for the two images exceeds two.

A point source at angular separation $\thetas > \thetaE$ produces only one image at
the angular position $\theta = \thetas + \thetaE$, with magnification
$(\thetaE+\thetas)/\thetas$, giving a value between 1 and 2. Thus if a source is a few Einstein radii away from the line of sight, it may still experience substantial magnification and differential magnification, and have observable signatures.

\subsection{Lensing probability}

For a background  elliptical galaxy at redshift $\zs$,
the lensing probability (optical depth) can be obtained as follows
\begin{eqnarray} \label{eq:lp1}
\tau (z_{\rm s})  &=& \int_0^{z_i}d\zd \int d \sigma
\,\Phi_{\sigma}(\zd,\sigma) \, {\sigma}_{\rm cr}(\sigma) \, {c\, dt \over d\zd},
\end{eqnarray}
where ${\sigma}_{\rm cr}$ is the cross-section for moderate
lensing, $t$ is the cosmic time, and $\Phi_{\sigma}(\zd,\sigma)$ is the lens velocity dispersion
function (e.g. Turner, Ostriker $\&$ Gott 1984,
Gott, Park $\&$ Lee 1989, Fukugita, Futamase $\&$ Kasai
1990). 

The lens velocity dispersion function  is modelled by a
modified Schechter function (\citealt{cho07}):
\begin{equation} \label{eq:lenvelfun}
\Phi_{\sigma}(\sigma) d\sigma = \Phi_{\sigma}^{\star} \left({\sigma \over
\sigma_\star}\right)^{\alpha_{\sigma}} \exp\left[-\left({\sigma \over
\sigma_\star}\right)^{\beta_{\sigma}}\right] {\beta_{\sigma} \over
\Gamma(\alpha_{\sigma} / \beta_{\sigma})} {d\sigma \over \sigma},
\end{equation}
where $\Phi_{\sigma}^{\star} = 8.0 \times 10^{-3} ~h^3
\rm~Mpc^{-3}$, $\alpha_{\sigma} = 2.32 \pm 0.10$, $\beta_{\sigma} =
2.67 \pm 0.07$, and $\sigma_\star = 161 \pm 5 ~\rm km s^{-1}$.
We limit the velocity dispersion to
the interval $70 \sim 400~\rm km s^{-1}$ (\citealt{loe03}). Most
massive early-type galaxies were already assembled at $z < 1$;
beyond this redshift, the density of early-type galaxies declines
significantly (e.g., \citealt{Renzini06}), and so we truncate the redshift of
early-type background galaxies at redshift 2. Below this truncation redshift,
we assume a constant comoving number
density of lenses, implying the physical volume density evolves as
$\Phi_{G,L}(z_d,\sigma) =
\Phi_{G,L}(\sigma)(1+z)^3$, where the factor $(1+z)^3$ accounts for
the expansion of the universe. Since most of the foreground (lensing) galaxies are below redshift 1, this assumption will not have much effects on our results.

\subsection{Elliptical galaxy luminosity function}

The luminosity function (LF) of early-type galaxies is modelled by the \citet{sch76} form
\begin{equation} \label{eq:lf1}
\Phi_{L}(L)dL = \Phi_{L}^\star \left({L \over
L_\star}\right)^{\alpha_{L}} \exp\left(-{L \over L_\star}\right) {dL \over
L_\star}.
\end{equation}
We adopt values from  \cite{cho07} that were obtained using a  
galaxy sub-sample of the SDSS Data Release 5 with
$M_\star - 5 \log_{10} h = -20.23 \pm 0.04$, $\alpha_L =
-0.527 \pm 0.043$, and $\Phi_L^\star= 0.71 \times 10^{-2}
h^{-3} \rm Mpc^{-3}$, appropriate for the SDSS $r$-band.

It is more convenient for our calculation here to use the luminosity
function with absolute magnitude $M$ rather than $L$. Eq. (\ref{eq:lf1}) can be rewritten as
\begin{equation} \label{eq:lf2}
\Phi_L(M)dM=0.4 \ln10 \times \Phi_L^\star
10^{-0.4 (M-M_\star)(\alpha_L+1)}  \exp(-10^{-0.4
(M-M_\star)})dM,
\end{equation}
where $M_\star$ is given below eq. (\ref{eq:lf1}).

Each survey has its own flux limit which corresponds to a
magnitude limit in a photometric band (e.g. the SDSS $r$ band),
denoted by $m_r$. 
The absolute limiting magnitude, $M_r$, of a galaxy at redshift $z$ and galactic
coordinates $(l,b)$, can be constructed from the apparent magnitude limit
$m_r$ as follows:
\begin{equation} \label{eq:am1}
M_r = m_r-DM(z)-K_r(z),
\end{equation}
where
$DM(z)$ is the distance modulus, and
$K_r(z)$ is the K-correction \citep{hog99}. Here we use the $K_r(z)$
given by \cite{cho07}, and we have ignored dust extinction by the Milky Way and the foreground galaxy.

The  lensing probability averaged over all source redshifts is given by (e.g., \citealt{wyi10})
\begin{equation} \label{eq:meanopticaldepth}
<\tau> = \frac{1}{N_{\rm G}} \sum_{i=1}^{N_{\rm G}} \tau(z_i) =
\frac{1}{N_{\rm G}} \int_0^{z_{s,\star}} dz\, \frac{dV_{\rm
C}}{dz}\, \Psi_{G,S}(M(\zs),\zs) \, \tau(\zs),
\end{equation}
where $N_{\rm G}$ is the total number of background elliptical
galaxies, $V_{\rm C}$ is the comoving volume, and 
$\Psi_{G,S}(M_r(\zs),\zs)$
is the comoving number density of elliptical galaxies brighter than
$M_r(\zs)$ at redshift $\zs$. $N_G$ can be calculated as
\begin{equation}
N_G=\int_0^{z_{s,\star}} dz\, \frac{dV_{\rm C}}{dz}\,
\Psi_{G,S}(M_r(\zs),\zs).
\end{equation}
and $\Psi_{G,S}(M_r(\zs),\zs)$ as
\begin{equation}
\Psi_{G,S}(M_r(\zs),\zs) = \int_{-\infty}^{M_r(\zs)} \Phi_{\rm L}(M) dM,
\end{equation}
where $\Phi_{\rm L}(M)$ is the background elliptical galaxy
luminosity function (see eq. \ref{eq:lf2}), and $M_r(\zs)$ is the
 absolute limiting magnitude $M_r$ at redshift $\zs$ corresponding to the magnitude limit $m_r$. To be conservative and for simplicity, we use the LSST single-visit depth limit $m_r=24.7$ in our optical depth calculation.

\subsection{Lens and source surface brightness profiles}

We model the background source (and the lens) as an elliptical galaxy with a de
Vaucouleurs surface brightness profile:
\begin{equation}
I(R) = I_{\rm e} \exp(-7.67((R/R_{\rm e})^{1/4}-1)),
\label{eq:deVaucouleurs}
\end{equation}
where $R$ is the two-dimensional radius, $R_{\rm e}$ is the
effective radius within which half of the light is enclosed and $I_{\rm e}$ is
the surface brightness at the effective radius. We quote the effective 
radius as the geometric mean of the major and minor axes, $R_{\rm
e}=\sqrt{a_{\rm e} b_{\rm e}}$. The total apparent magnitude $m_{\rm
T}$, average magnitude within effective radius $\langle \mu \rangle_{\rm e}$ and
effective radius $R_{\rm e}$ can be related by \citep{sco98} 
\begin{equation}
m_{\rm T} = -2.5\rm{log_{10}} (L_{\rm T}) - 48.6 = \langle \mu\rangle_{\rm e} -
1.9954-5\rm log_{10}(R_{\rm e}), \label{eq:totalmag}
\end{equation}
where $L_{\rm T}$ is the total flux of the elliptical galaxy and
$R_{\rm e}$ is measured in arcsec.

\subsection{Simulating images of early-type galaxies}
\label{sec:sourceinfo}

In an image of an early-type galaxy, in addition to the signal of
the source, we also have the Poisson noise in the sky background, readout
noise of the CCDs and the dark current. The signal-to-noise ratio is
given by \citep{mcm08}
\begin{equation} \label{eq:signalNoiseRatio}
\frac{S}{N} = \frac{N_{\rm S} t}{\sqrt{N_{\rm S}t + n(N_{\rm sky}t +
N_{\rm DC}t + N_{\rm R}^2)}},
\end{equation}
where $N_{\rm S}$ is the number of photoelectrons from the source
per unit time, $N_{\rm sky}$ is the intensity of sky background in
photoelectrons per pixel per unit time, $N_{\rm DC}$ is the dark
current in electrons per pixel per unit time, $N_{\rm R}$ is the
readout noise in electrons per pixel, $n$ is the number of pixels
covered by an image of an early-type galaxy and $t$ is the exposure time.

To proceed further, we need to adopt a concrete case for the
parameters in eq. (\ref{eq:signalNoiseRatio}). Here we restrict 
ourselves to LSST. The telescope will be sited
in Cerro Pach\'on, Chile, with excellent median free-air seeing of
$0.^{\prime\prime}7\,$ in the $r$ band. Correspondingly, the pixel
scale is chosen to be $0.^{\prime\prime}2$.  The telescope will
repeatedly survey the sky covering an area of $20000$ square
degrees with a cadence of $\approx 7$ days.
The integration time ($t$) for each individual frame is 15 seconds. We
assume that the readout noise is the same for all exposures. For the
Poisson noise in the sky background, readout noise of the CCDs and
the dark current, we use the mean values given by the LSST Exposure Time
Calculator\footnote{http://dls.physics.ucdavis.edu:8080/etc4\_3work/servlets/LsstEtc.html}. 
Using these numbers, we verified that the photon numbers we obtain differ from those given by the LSST Exposure Time Calculator by less than $3\%$. 

\begin{figure}
\begin{center}
\begin{minipage}[t]{1.0\hsize}
 \includegraphics[width=\hsize]{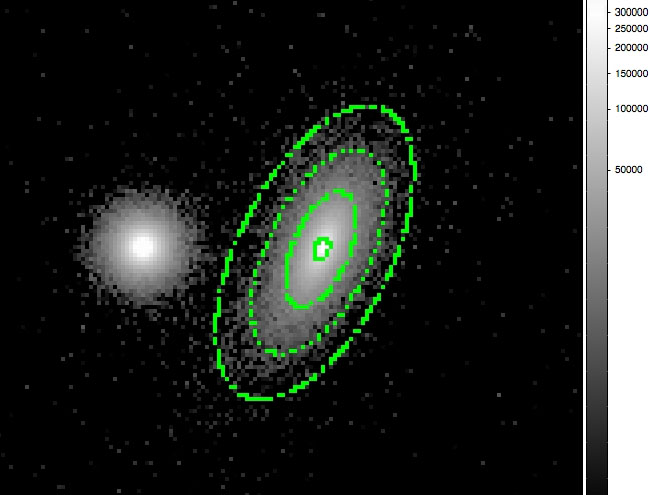}
\end{minipage}
\end{center}
\caption{A simulated lensed image of a background elliptical galaxy,
with a scale of $0.^{\prime\prime}2$ per pixel. The foreground
  lens is at redshift 0.3 and the background elliptical galaxy (on the
  right) is at redshift 0.6. The size of the image is
$30.^{\prime\prime}0$ by $25.^{\prime\prime}6$.  The distance
  between the two galaxy centres is $8.2$ arcsec.
The ellipses are obtained by running the IRAF task {\tt ELLIPSE} (see \S
\ref{sec:fitting}). Notice the small changes in the position angle and
twists in the isophotes.}   
\label{fig:case}
\end{figure}

Fig.\,\ref{fig:case} shows an example of a simulated moderately 
lensed background elliptical galaxy over-plotted with the best-fit ellipses from
the IRAF task {\tt ELLIPSE}  (for more details, see \S
\ref{sec:fitting}). Notice that we have not convolved the image with seeing.

To simulate a population of early-type
  galaxies, we use the method given in Appendix A of \citet{ber03a}. Briefly, each galaxy is randomly assigned an absolute magnitude,
  $M_{\rm r}$, effective radius $R_{\rm e}$ and velocity dispersion
  $\sigma$ ($V$ in the notation of \citealt{ber03a}), appropriate at
  redshift $z=0$. This procedure ensures the generated
   galaxies satisfy the observed correlations among  observed properties,
    including the fundamental plane as given by \citet{ber03b} 
  \begin{equation}
  \log_{10} R_{\rm e} = \log_{10} \sigma+0.2(\langle \mu\rangle_{\rm e}-20.09).
  \label{eq:fund}
  \end{equation} 
 We assign a typical error of $0.03$\,dex in $\log\sigma$ and $0.02$\,dex in $\log_{10} R_{\rm e}$. The scatter in the fundamental plane is quite small, on the order of
 $\approx 0.08$, and may be mostly due to measurement errors.  
  
The \citet{kor77} relation between the surface brightness and effective radius is given by (\citealt{ber03a}, as quoted in \citealt{ooh09})
  \begin{equation}
 \langle \mu\rangle_{\rm e}  = 2.04 \, \log_{10} R_{\rm e} + 18.7,
  \label{eq:fund2}
  \end{equation} 
  where $\langle \mu\rangle_{\rm e}$ is the mean surface brightness (in mag/arcsec$^2$) within the effective radius ($R_{\rm e}$, in units of kpc). The scatter in this relation is substantially larger than that in the fundamental plane (see \S\ref{sec:kormendy}).
 
 For LSST, the mean redshift of the galaxy sample is about $0.8$ (see the LSST science book 
for details). However, to be specific and somewhat conservative, we use a lens redshift of 0.3 and a source redshift of 0.6. In reality, galaxy pairs will have a redshift distribution,  but we can in principle bin the data and study the quantities we are interested, and so this simplification does not change the results of the paper. Notice that for the optical depth calculation, we still assume the lenses and sources have a redshift distribution, as described in \S\ref{sec:model}.

 For each random galaxy generated for redshift 0, we shift it to redshift $z$ by taking into account its evolution \citep{ber03a}
 \begin{equation}
 M_r(z) = M_r(z=0)-Q z,
\label{eq:Mag_simu}
 \end{equation}
 where $Q=0.85$ for the $r$ band.
  
 We assume a total exposure time of 6000 seconds (i.e., 400 exposures
 of 15 seconds). We further select only galaxies which satisfy 
  $R_{\rm e}>0.^{\prime\prime}7$, as a way to crudely account for the
 effects of seeing. In total, we generate 11623 pairs of foreground and
 background galaxies which satisfy these conditions.
 The ellipticity of the background galaxy is
 randomly drawn from the axis ratio distribution as given in
 \cite{cho07} (see their Fig.13). The lensed background galaxy images
 are obtained using the lens equation. The following analyses are
 based on this sample, although we apply a further cut in the isophote shape analysis (see below). 
 
 \section{Results}
\label{sec:result}

\subsection{Optical depth}
\label{sec:opticalDepth}

\begin{figure}
\begin{center}
\begin{minipage}[t]{1.0\hsize}
\includegraphics[width=\hsize]{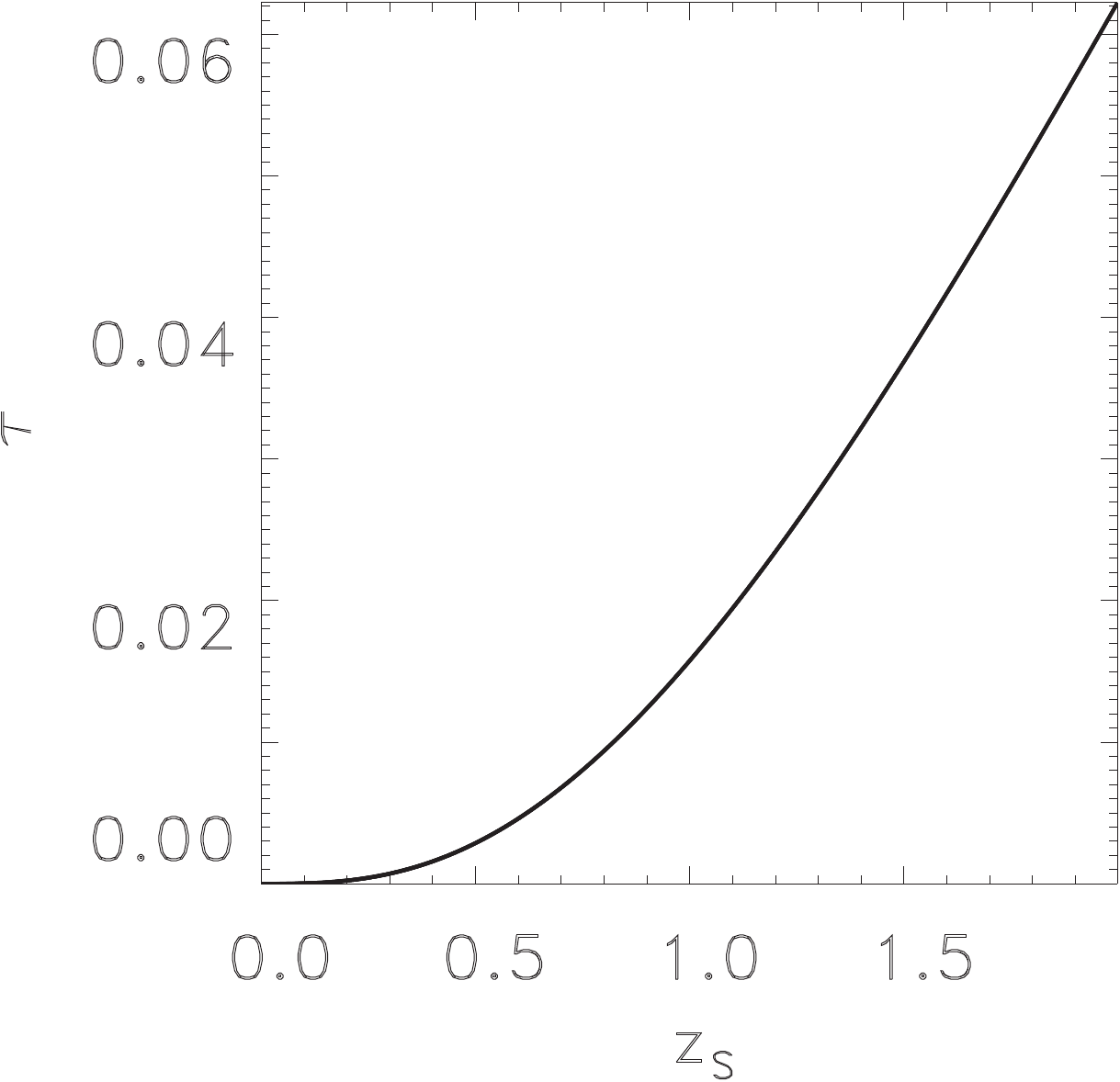}
\end{minipage}
\end{center}
\caption{The optical depth, $\tau$, as a function of the redshift,
$z_{\rm s}$. The cross-section for moderate lensing is taken as the
  area between $1\thetaE$ to $10\thetaE$. 
} 
\label{fig:redshift_opticaldepth}
\end{figure}

In Fig. \ref{fig:redshift_opticaldepth}, we present the optical
depth as a function of redshift. In this exercise, we take the moderate lensing cross-section
for each galaxy as $\thetaE$ to $10\thetaE$ (corresponding to
$\mu = $ 2 and 1.1).  Not surprisingly the optical depth increases with increasing redshift, reflecting the fact that the 
number of intervening galaxies increases for more distant background sources.
The optical depth is about 0.0029 at $z = 0.5$, 0.016 at $z = 1.0$,
and 0.062 at $z=2.0$. The  optical depth of averaged over all the background
sources is about 0.0254. This probability
will scale linearly with the
cross-section we adopt. For example, if the we take the cross-section from
$2\thetaE$ to $5\thetaE$ (corresponding to $\mu=1.5$ and 1.2)  then the probability will be reduced by a factor of 
roughly 5 to about 0.005. The precise choice of cross-section will depend on the depth, seeing etc. of observations.

We have also calculated the magnification bias (\citealt{tur84}),
which can be large for multiply-imaged quasars or galaxies. In our
case, we find the magnification bias to be modest, only about 1.16. The
magnification bias is small because the magnification in our
case is by definition modest.  

\begin{figure}
\begin{center}
\begin{minipage}[t]{1.0\hsize}
\includegraphics[width=\hsize]{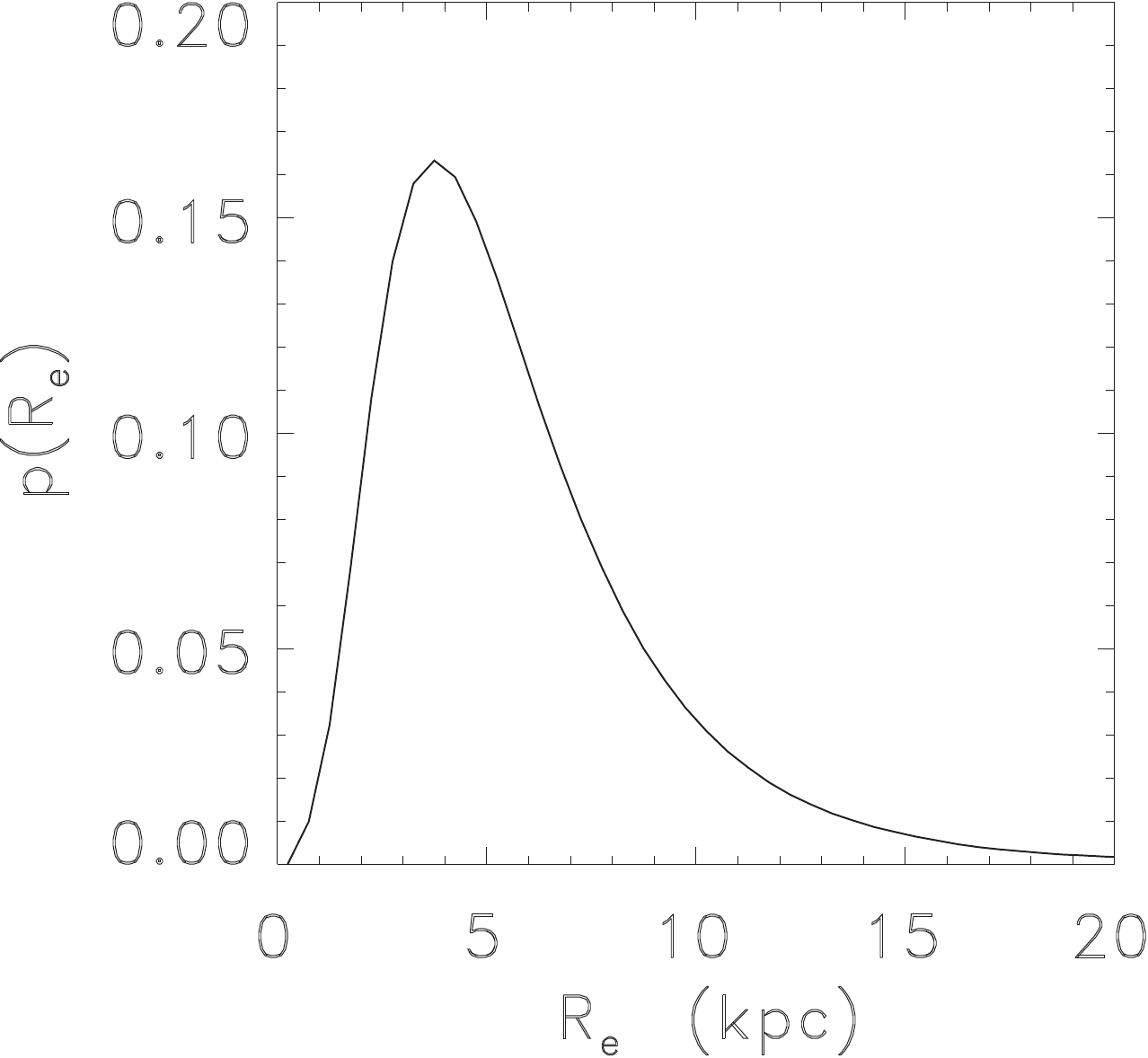}
\end{minipage}
\end{center}
\caption{The probability density distribution of the effective radii of the
background elliptical galaxies, all at redshift 0.6.
At this redshift, 1 arcsec corresponds to $4.7 h^{-1}$\,kpc.
}
\label{fig:Re_distribution0_1log}
\end{figure}

As we have mentioned in \S\ref{sec:sourceinfo}, the redshift of
the background sources is truncated at $z=2.0$. With this
limitation, the surface number density of early-type galaxies is
about 10 per square arcmin. For LSST, it will carry out a survey of
$20000\,\rm deg^2$ of the sky, and about 700 million elliptical
galaxies will be detected. In Fig. \ref{fig:Re_distribution0_1log}, we
present the distribution of the effective radius of the background elliptical galaxy sources.
The number of elliptical galaxies drops steeply for larger
effective radii. About $54.1\%$ of the background sources
have effective radii larger than $0.^{\prime\prime}7$, 
the median free-air seeing at the site of LSST. Under this
criterion, we estimate that there are about $9.5$ million
moderate galaxy lensing cases that will be observed by LSST to
the single-visit depth ($m_r=24.7$ mag). 

\begin{figure}
\begin{center}
\begin{minipage}[t]{1.0\hsize}
\includegraphics[width=\hsize]{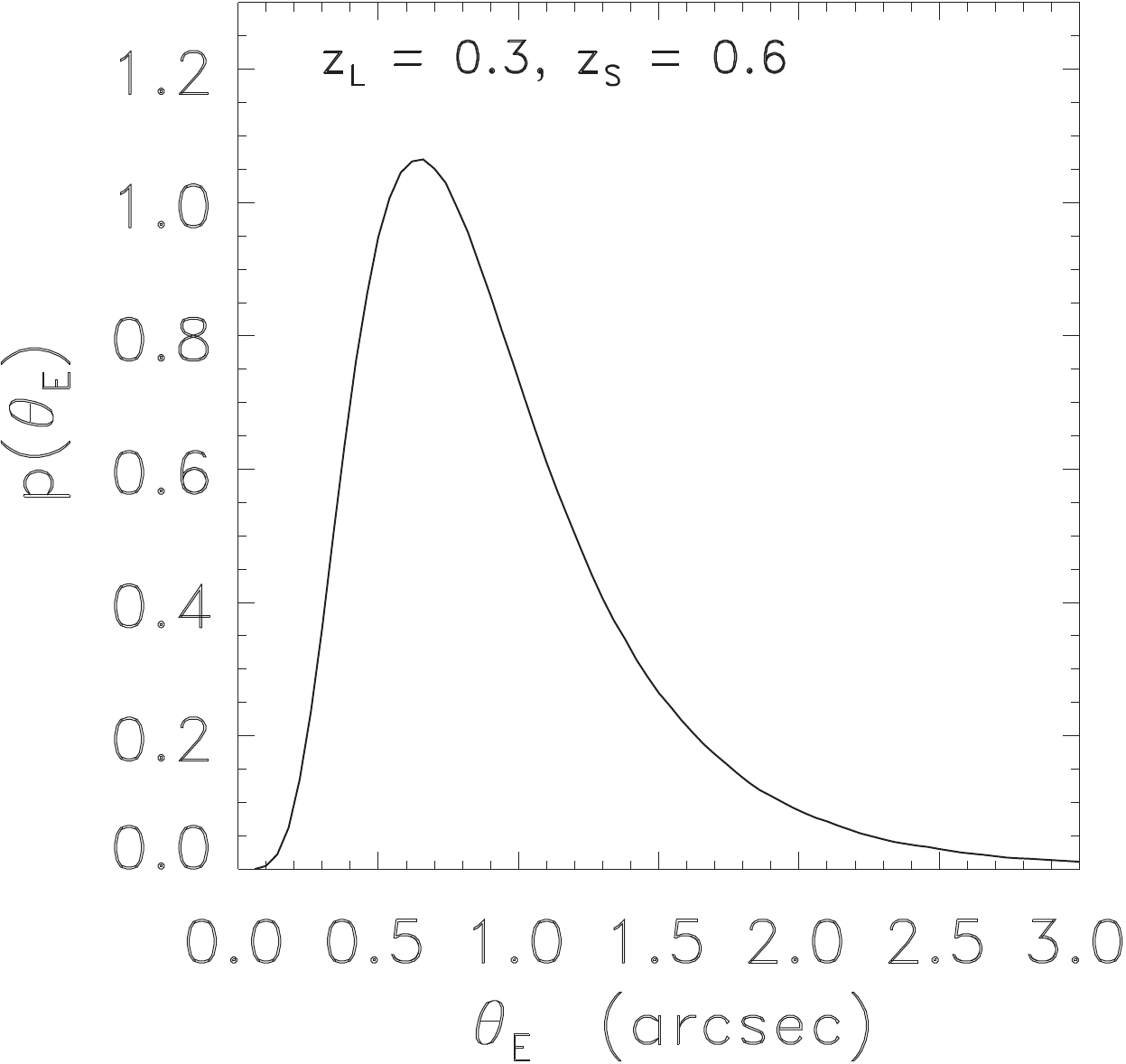}
\end{minipage}
\end{center}
\caption{The probability density distribution of the angular Einstein
  radius \,$\thetaE$\, 
of the simulated lensing cases.
} \label{fig:thetaE_distribution}
\end{figure}

The predicted angular Einstein radius is shown in
Fig. \ref{fig:thetaE_distribution}. As can be seen, the distribution
peaks around 0.8 arcsec and has an extended tail out to larger separations.
The distribution is similar to the observed one from 
the CLASS survey, even though the source and lens distributions there
are somewhat different (see Fig. 11 in \citealt{bro03}).

\subsection{Isophotal distortions}
\label{sec:fitting}

Surface photometry is performed on the simulated lensed images of the
background elliptical galaxy by utilizing the IRAF task {\tt ELLIPSE}
(\citealt{Jedrzejewski87}). When running the {\tt ELLIPSE} on the lensed
images, the geometric centre, elllipticy and position angle are
all allowed to vary freely, with a logarithmic step of 0.05 in the semi-major
axis. The isophotes of the lensed images can be measured in different
ways. Here we present the output of the task ELLIPSE. A
detailed description of these measurements can be found in 
\citet{Hao06}, we only repeat the essentials here.  The
  intensity along the ellipse is expanded in Fourier series
\begin{equation}
 I (\theta) = I_0 + \sum (A_n \cos n \theta + B_n \sin n \theta),
 \label{eq:Itheta}
\end{equation}
where $I_0$ is the intensity averaged over the ellipse, and $A_n$ and
$B_n$ are  the higher order Fourier coefficients. If an isophote is a
perfect ellipse, then all the coefficients, $(A_n, B_n), n=1,
\cdot\cdot\cdot, \infty$ will be exactly zero. We will use two quantities
$a_n$ and $b_n$ extensively, which are related to the
 {\tt ELLIPSE}
outputs $A_n$ and $B_n$  by
\begin{equation}
\frac{a_n}{a} = \frac{A_n}{\gamma a},
\end{equation}
where $\gamma = dI/dR$ is the local radial intensity gradient, and $a$ is the semi-major axis length.

\begin{figure}
\begin{center}
\begin{minipage}[t]{1.0\hsize}
\includegraphics[width=\hsize]{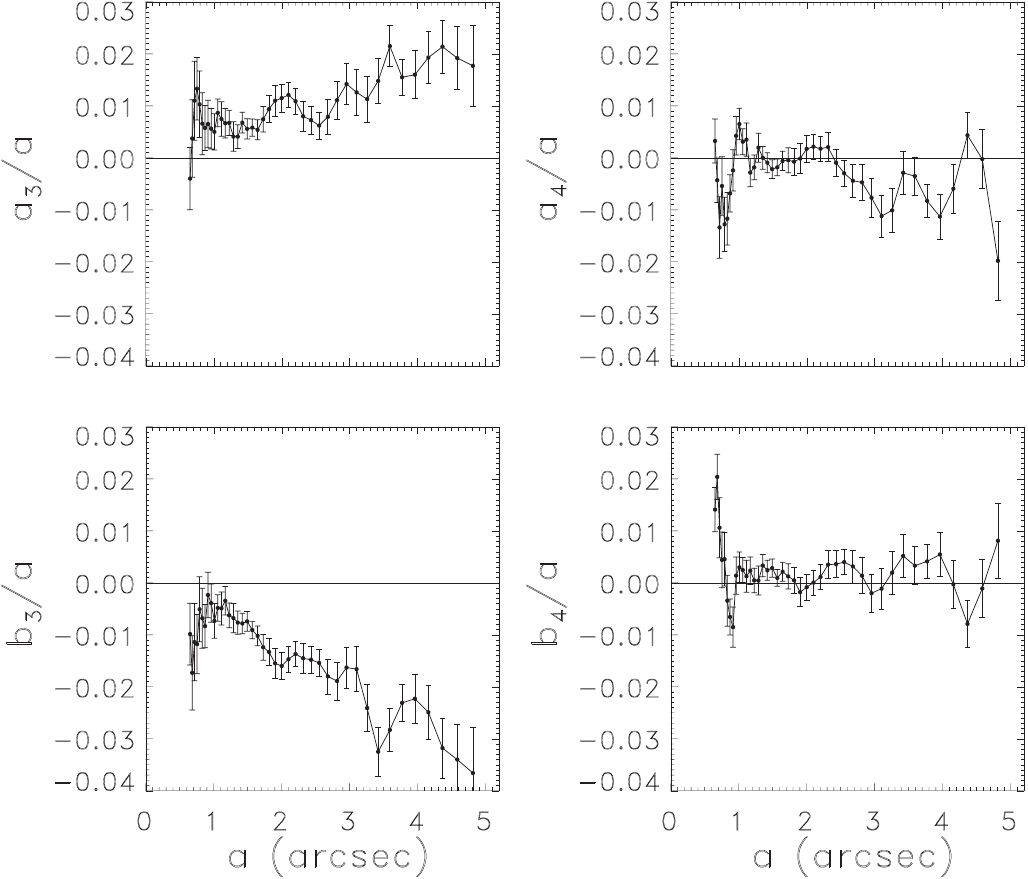}
\end{minipage}
\end{center}
\caption{Coefficients of the Fourier series $a_3/a,\, a_4/a,\,
b_3/a$ and $b_4/a$ obtained by the IRAF task ELLIPSE for the galaxy shown in 
Fig.\,\ref{fig:case}. The horizontal line in each panel indicates the
value for a perfect ellipse. 
}
\label{fig:case_a3a4b3b4}
\end{figure}

For the galaxy shown  in Fig.\,\ref{fig:case}, the coefficients of
the Fourier series $a_3/a,\, a_4/a,\, b_3/a$ and $b_4/a$
as a function of the semi-major axis radius are shown in Fig.\,\ref{fig:case_a3a4b3b4}. The horizontal line in each panel is the value for a perfect
ellipse. Notice that both $a_3/a$ and $b_3/a$ show 
significant deviations from the isophotes of a perfect ellipse.

\begin{figure}
\begin{center}
\begin{minipage}[t]{1.0\hsize}
\includegraphics[width=\hsize]{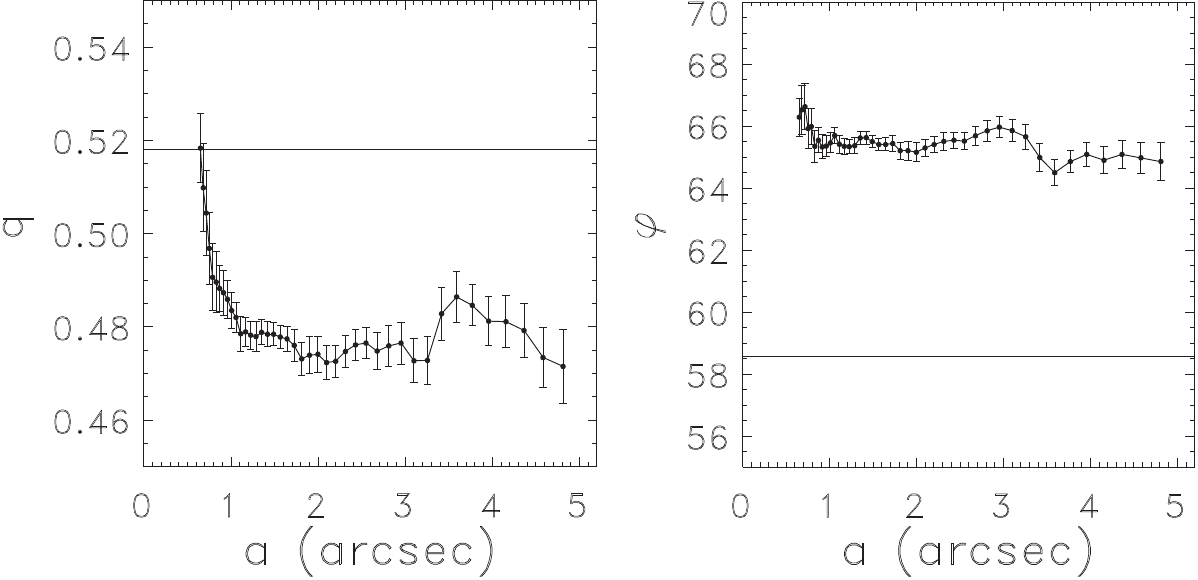}
\end{minipage}
\end{center}
\caption{The probability density distributions of the apparent isophotal axis ratio ($q$),
position angle ($\varphi$, in units of degrees) for the galaxy shown in
Fig. \ref{fig:case}. }
\label{fig:case_qphi}
\end{figure}

For the same galaxy, we present the distribution of
the apparent isophotal axis ratio $q$,  position angle $\varphi$ in
Fig.\,\ref{fig:case_qphi},. The horizontal line presents the
ellipticity (position angle) before lensing.
There are some small but systematic changes in these two
quantities of a few percent for the example shown here.
The $a_4/a$ and $b_4/a$ parameters describe whether an elliptical
galaxy is disky or boxy, but we did not incorporate these into our
image simulations. Instead we choose to focus on the $a_3/a$ and
$b_3/a$ parameters.
 
\begin{figure}
\begin{center}
\begin{minipage}[t]{1.0\hsize}
\includegraphics[width=\hsize]{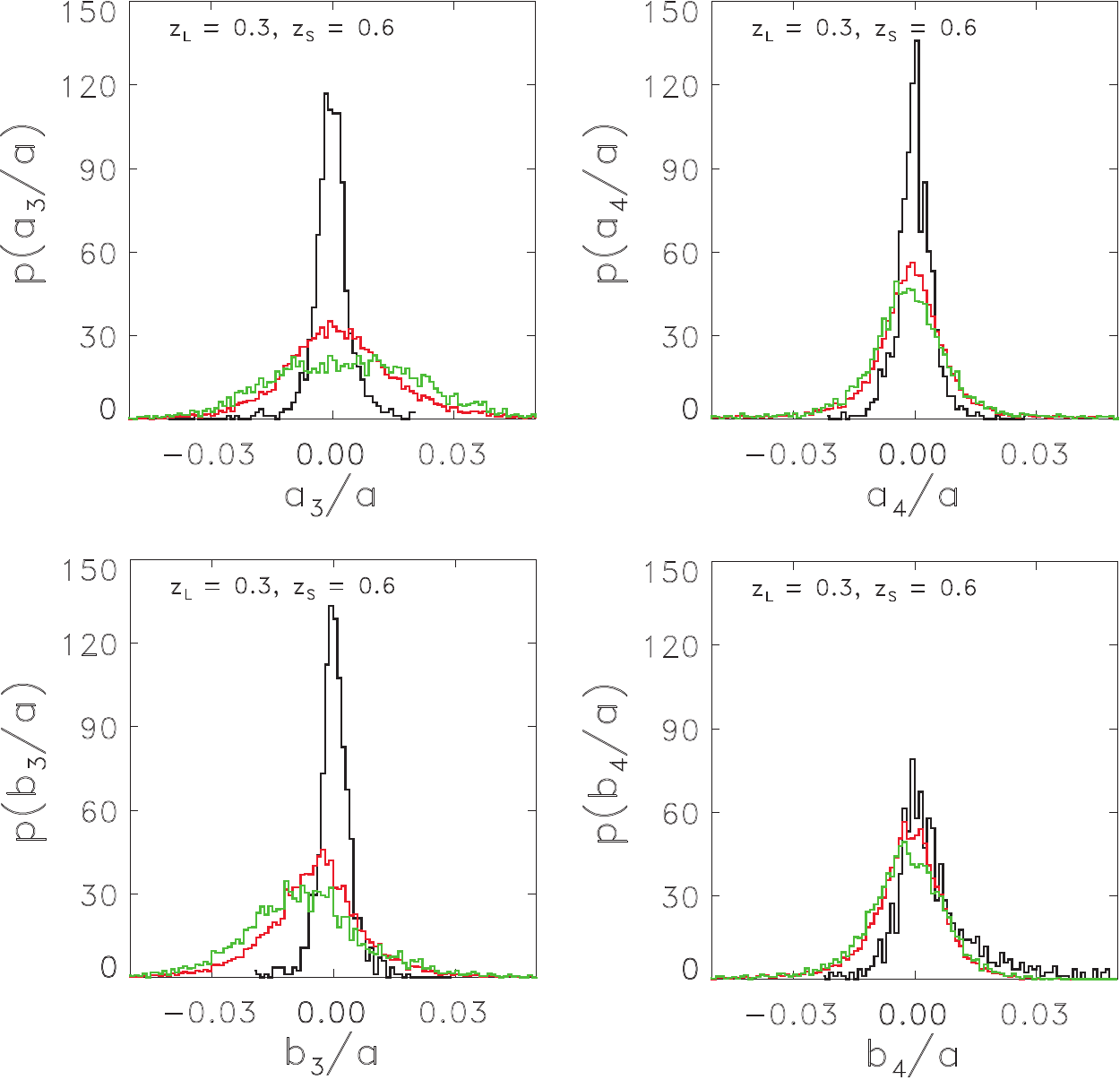}
\end{minipage}
\end{center}
\caption{The probability density distributions of the coefficients of the Fourier series
  $a_3/a,\, a_4/a,\, b_3/a$ and $b_4/a$ obtained by the IRAF task
  {\tt ELLIPSE}.  The black solid histogram in each panel shows the distribution of the
  coefficients for a nearby sample of early-type galaxies from  \citet{Hao06}. 
  The green line is for moderately lensed
  galaxy pairs within separation between 2 to 6 arcsec while the red solid line is for galaxies from 2 to 10 arcsec. 
  }
\label{fig:a3a4b3b4_2}
\end{figure}

In Fig.\ref{fig:a3a4b3b4_2}, 
we present the distribution of the coefficients of the Fourier series
$a_3/a,\, a_4/a,\, b_3/a$ and 
$b_4/a$ for  cases where the separations between the foreground galaxy and
background galaxy are between  $\thetaE + R_{\rm e, s} \approx
  2.^{\prime\prime}0$ and
$10.^{\prime\prime}0$, where $R_{\rm e, s}$ is the effective radius for the source.
For each galaxy, these coefficients are the mean value
  within the region between $2r_s$ and $1.5R_{50}$, where $r_s
  \approx 0.^{\prime\prime}7$ is the median seeing radius and $R_{50}$ is the Petrosian
  half-light radius. For SDSS, $R_{50} = 0.71 R_{\rm e}$ for early-type
  galaxies described by the $R^{1/4}$ law \citep{Hao06, Graham05}. 
 In the same plot, these distributions for
separations between $2.^{\prime\prime}0$ and
$6.^{\prime\prime}0$ and between $2.^{\prime\prime}0$ and
$10.^{\prime\prime}0$ are also shown. As expected, for background sources with smaller separations from the lens
  centre, the distributions of the coefficients are broader
   because the lensing effects are stronger in such cases.
  Notice that in the above analysis, we require $2 r_s > 1.5 R_{50}$, this condition reduces the pair of galaxies from 11623 to 9094.
      
\begin{figure}
\begin{center}
\begin{minipage}[t]{1.0\hsize}
\includegraphics[width=\hsize]{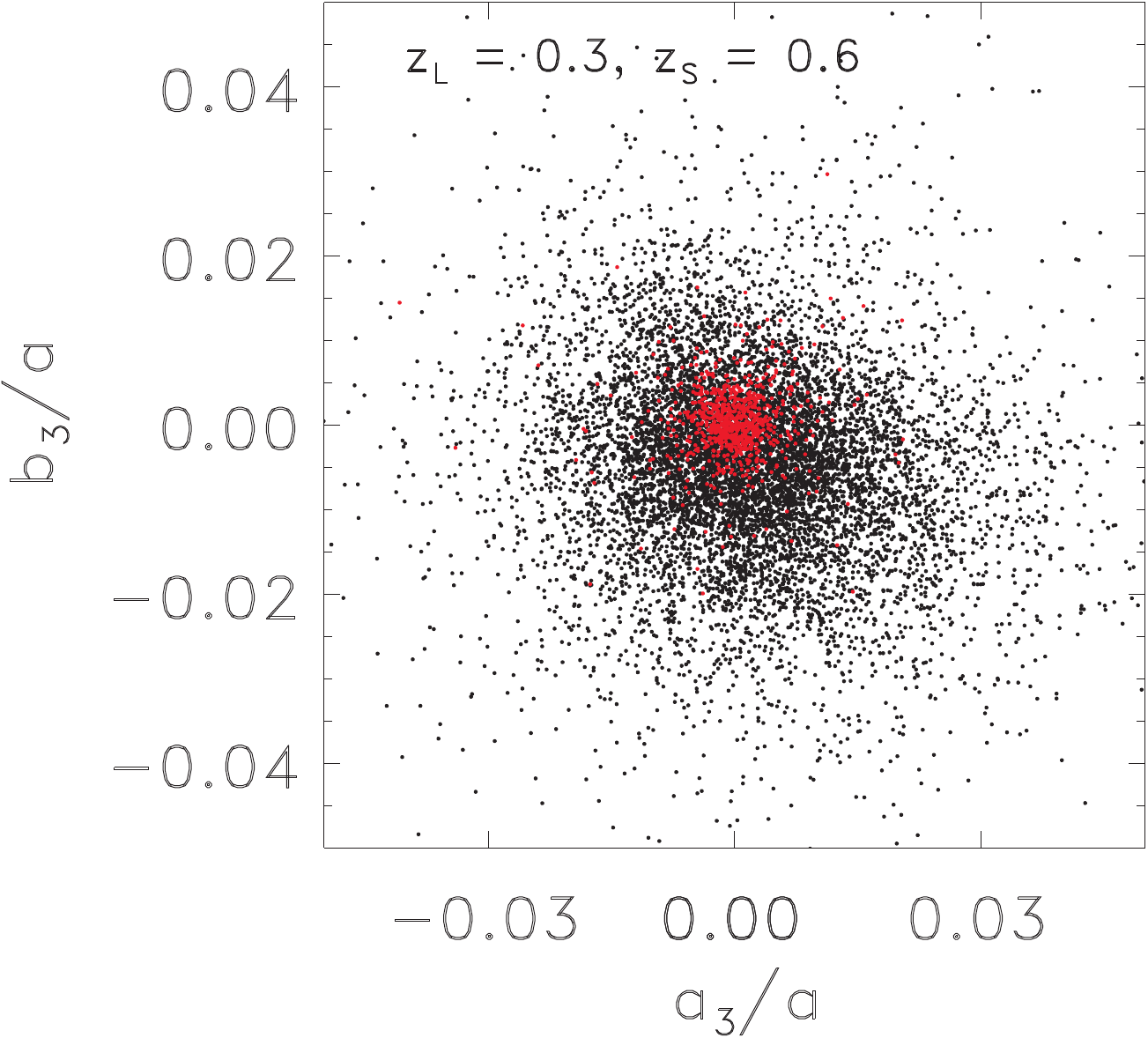}
\end{minipage}
\end{center}
\caption{The scattered plot for the simulated galaxy pairs in the
  plane $a_3/a$ vs. $b_3/a$. The red dots are for unlensed galaxies
  while the black ones are for lensed ones. In total, there are
  9094 pairs of galaxies. }
\label{fig:a3_b3}
\end{figure}

Both the $a_3/a$ and $b_3/a$ distributions show
large deviations from those of normal elliptical
galaxies as studied by \citet{Hao06}. This is further illustrated in Fig. \ref{fig:a3_b3}, here we plot the two-dimensional distributions of these two quantities rather than as histograms from projection.
We can see that the unlensed galaxies are mostly at the centre, while the
lensed galaxies clearly show much larger scatters than the unlensed ones. 
Thus they can be used as effective indicators whether
moderate gravitational lensing has occurred in a pair. 

\subsection{Fundamental plane}
\label{sec:fundamentalplane}

Gravitational lensing preserves the surface brightness of the source
because of Liouville's theorem, but changes the apparent solid angle
of a source. Following this, a natural expectation is that the fundamental
plane of  background elliptical sources should be affected by moderate
gravitational lensing. 

To obtain the lensed and unlensed effective radii and surface brightness, $R_{\rm e, L}$ and $R_{\rm e}$, we fit
the surface brightness profile of the lensed images of background
elliptical galaxies with the $R^{1/4}$ law. In this model, we fix the
centre of the galaxy, after which there are four remaining free
parameters, $\{I_{\rm e},\, R_{\rm e},\, \varphi,\, q\}$. The $\chi^2$
function is given by  
\begin{equation}
\chi^2(I_e,R_{\rm e},\varphi,q) = \sum_{i=1}^{N_{p}}
\frac{({\cal I}_i-{\cal I}_{i,0})^2}{\sigma_{i,0}^2}. \label{eq:inverseFitSimplex}
\end{equation}
where $N_{\rm
p}$ is the number of pixels of the lensed image. ${\cal I}_{i,0}$ is the
photon number in the $i^{\rm th}$ pixel, $\sigma_{i,0}$ is the Poisson noise $\sigma_{i,0} = \sqrt{{\cal I}_{i,0}}$. The degree of freedom is $N_{\rm dof} = N_{\rm
  p}-4$. We use the routine gsl\_multimin\_fminimizer\_nmsimplex in the GNU
Scientific Library\footnote{ http://www.gnu.org/software/gsl/} to 
find the $\chi^2$ minimum and obtain the effective
radius; the mean surface brightness is then obtained by averaging
the surface brightness within $R_{\rm e}$.

\begin{figure}
\begin{center}
\begin{minipage}[t]{1.0\hsize}
\includegraphics[width=\hsize]{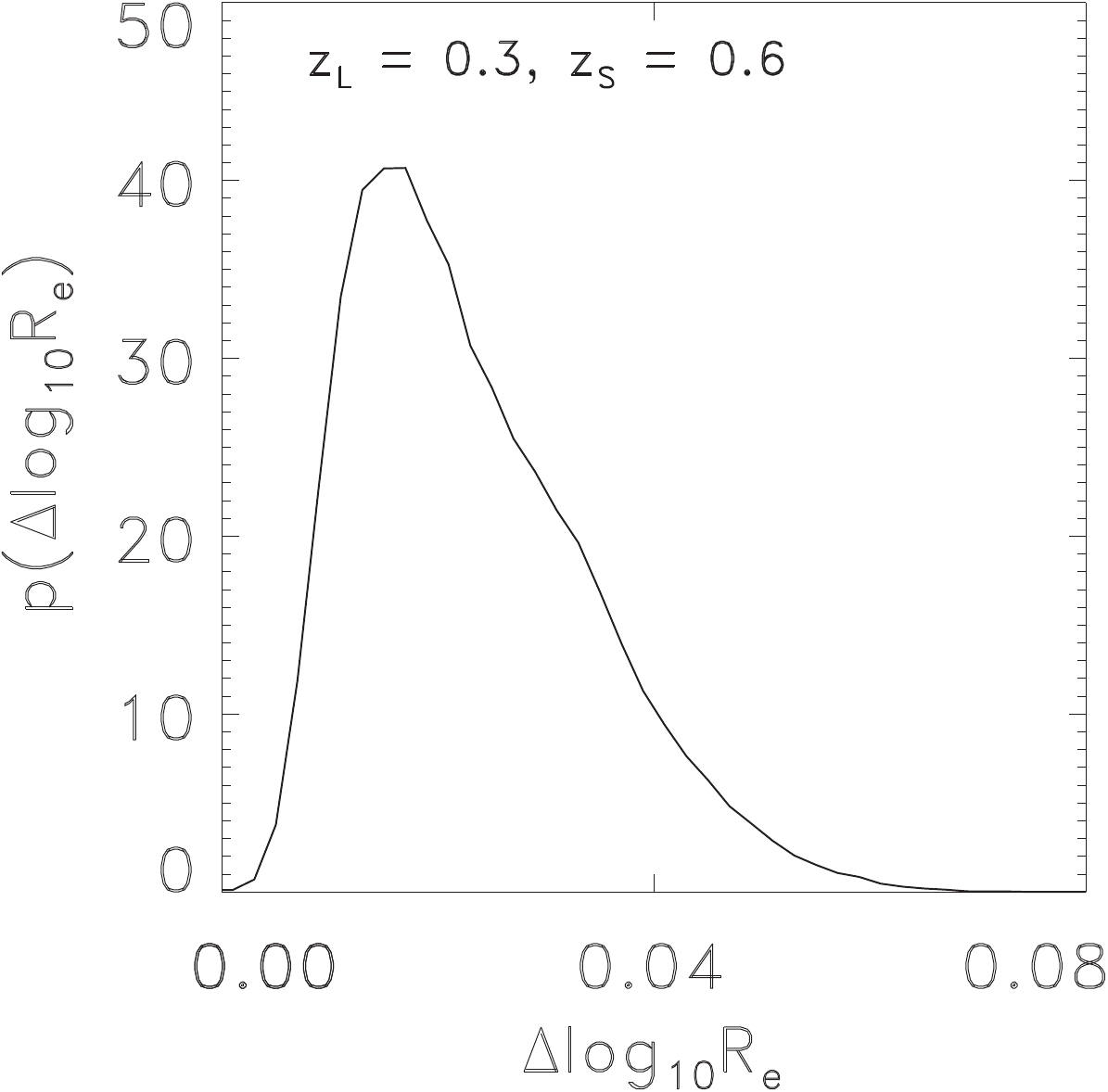}
\end{minipage}
\end{center}
\caption{The probability density distribution of differences in effective radii between the
  lensed and unlensed sources in our simulated
  sample. Notice the systematic shift to a larger effective
  radius. 
  } \label{fig:Remag_distribution}   
\end{figure}

Fig.\,\ref{fig:Remag_distribution} shows the distribution of
differences in effective radii between the lensed and unlensed sources,
defined as $\Delta \log_{10}
yR_{\rm e} =  \log_{10} R_{\rm e,L} - \log_{10} R_{\rm  e}$.
The mean and median values of $\Delta \rm 
log_{10} R_{\rm e}$ are 0.03 and 0.03 respectively. These values will depend on the
separation range we take for the galaxy pair, here we have
  selected pairs with separations between $\sim$ 2 to 10 arcsec.
  
\begin{figure}
\begin{center}
\begin{minipage}[t]{1.0\hsize}
\includegraphics[width=\hsize]{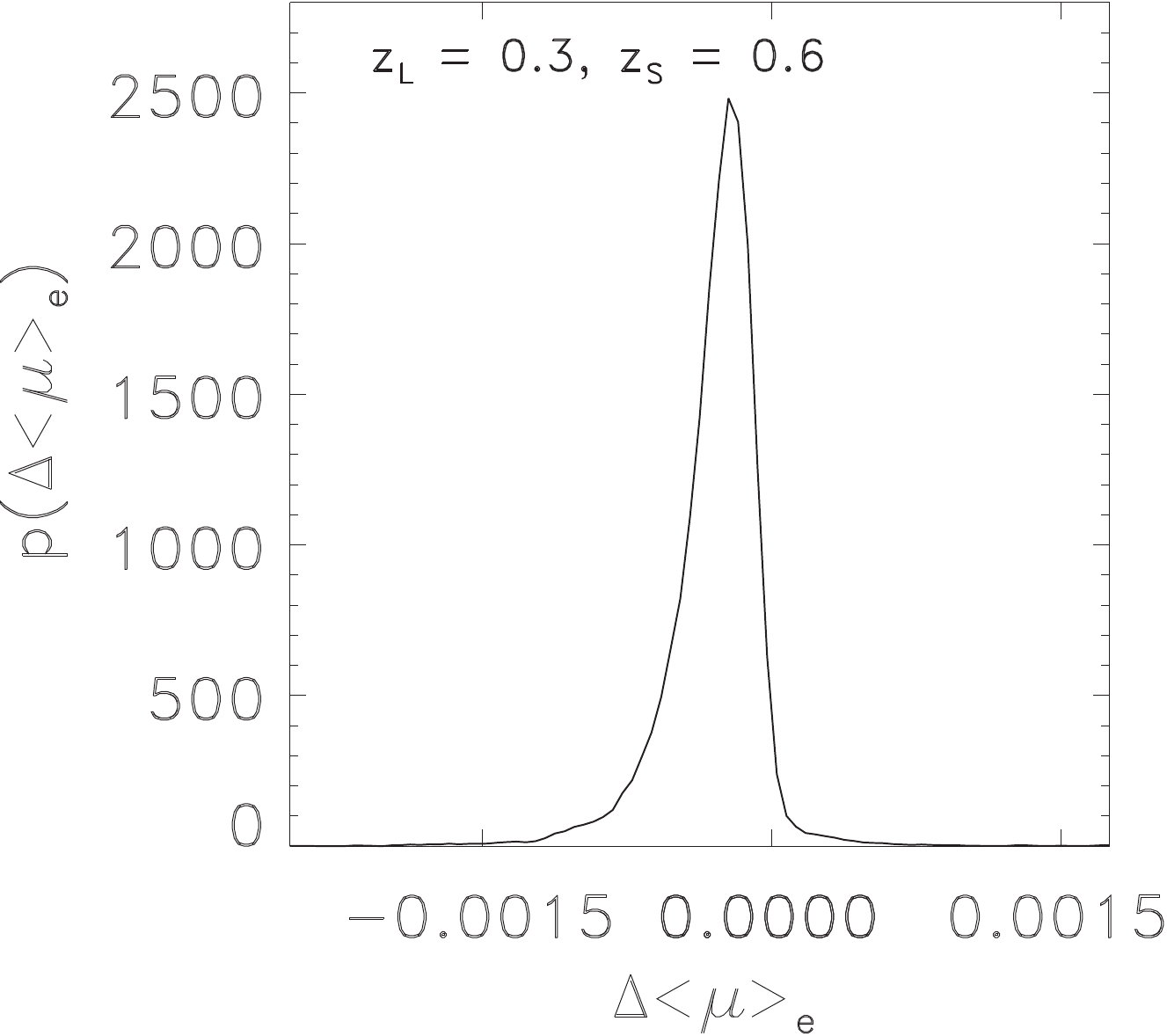}
\end{minipage}
\end{center}
\caption{The probability density distribution of differences in mean surface brightness within
  effective radius between the lensed and unlensed sources in our
  simulated sample. 
 } \label{fig:mumag_distribution}   
\end{figure}

Similarly, Fig.\,\ref{fig:mumag_distribution} shows the distribution of
differences in the mean surface brightnesses within the effective radius between
the lensed and unlensed sources, defined as
 $\Delta \langle \mu  \rangle_{\rm e}= \langle \mu\rangle_{\rm e, L} - \langle\mu\rangle_{\rm e}$, where $\langle\mu\rangle_{\rm e,L}$ and $\langle\mu\rangle_{\rm e}$ are the
mean surface brightness within the lensed and unlensed effective radius of the
lensed images.  The mean and median differences
are $-0.0003$ and $-0.0003$, which are much
smaller than those for the effective radius, i.e.,
gravitational lensing does not affect the mean surface 
brightness within the effective radius. 

The fundamental plane of the lensed
elliptical galaxies is presented in Fig. \ref{fig:Fundamental_Plane}.
 The black solid line shows the fundamental plane used to produce
the background elliptical galaxies (black dots) while the red symbols present the
lensed elliptical galaxies. The best regression line for the lensed sample is shown as the dashed red line. On average an upward offset of about 0.03 is introduced by gravitational lensing.

\subsection{Kormendy relation}
\label{sec:kormendy}

\begin{figure}
\begin{center}
\begin{minipage}[t]{1.0\hsize}
\includegraphics[width=\hsize]{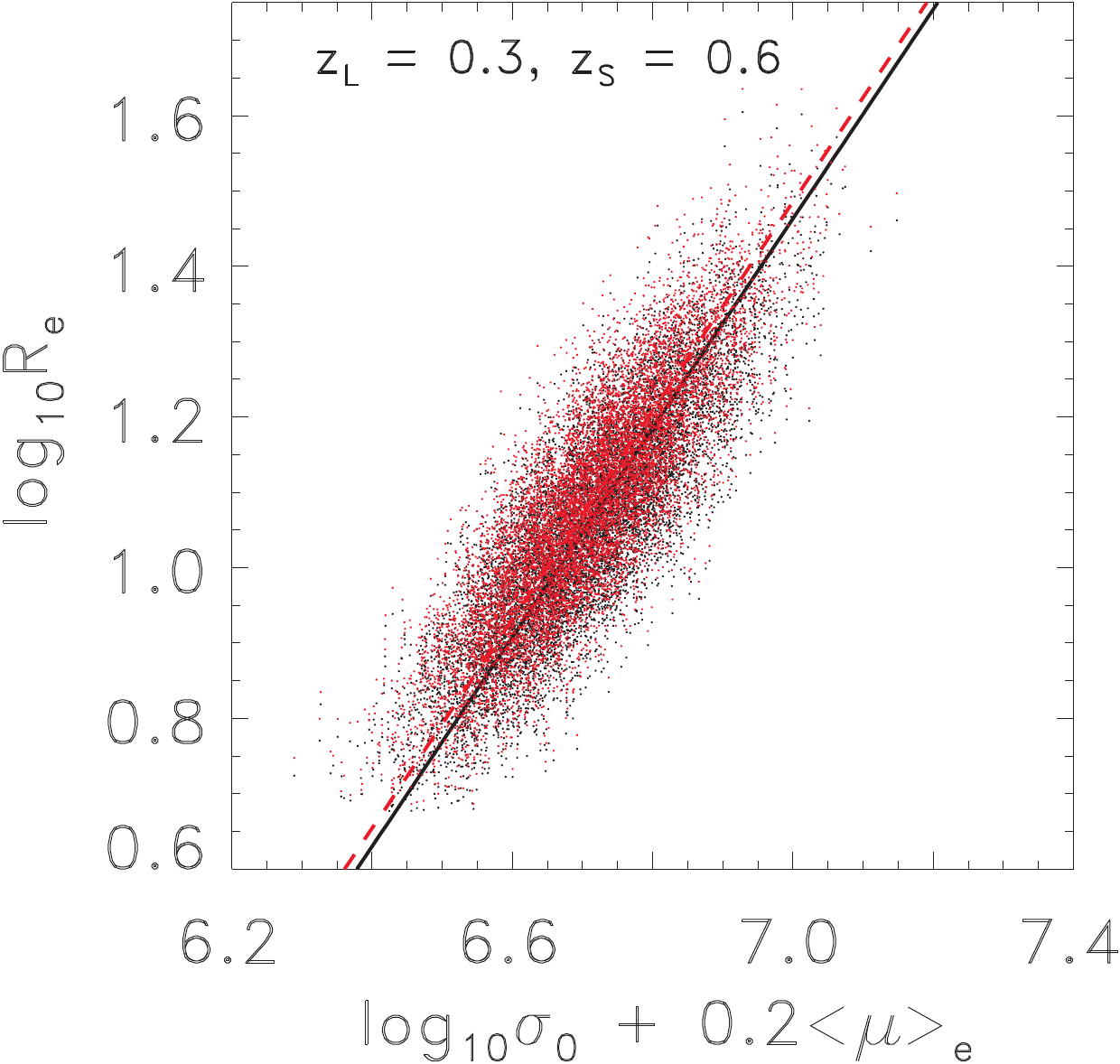}
\end{minipage}
\end{center}
\caption{Fundamental plane of the lensed elliptical galaxies. The
black solid line shows the fundamental plane for the
background elliptical galaxies. The red dots are for the
lensed elliptical galaxies with the best regression line shows as the red dashed line. The lens galaxies are 
at redshift $z_{\rm l} = 0.3$ and the source galaxies at $z_{\rm s}=0.6$.}
 \label{fig:Fundamental_Plane}
\end{figure}

\begin{figure}
\begin{center}
\begin{minipage}[t]{1.0\hsize}
\includegraphics[width=\hsize]{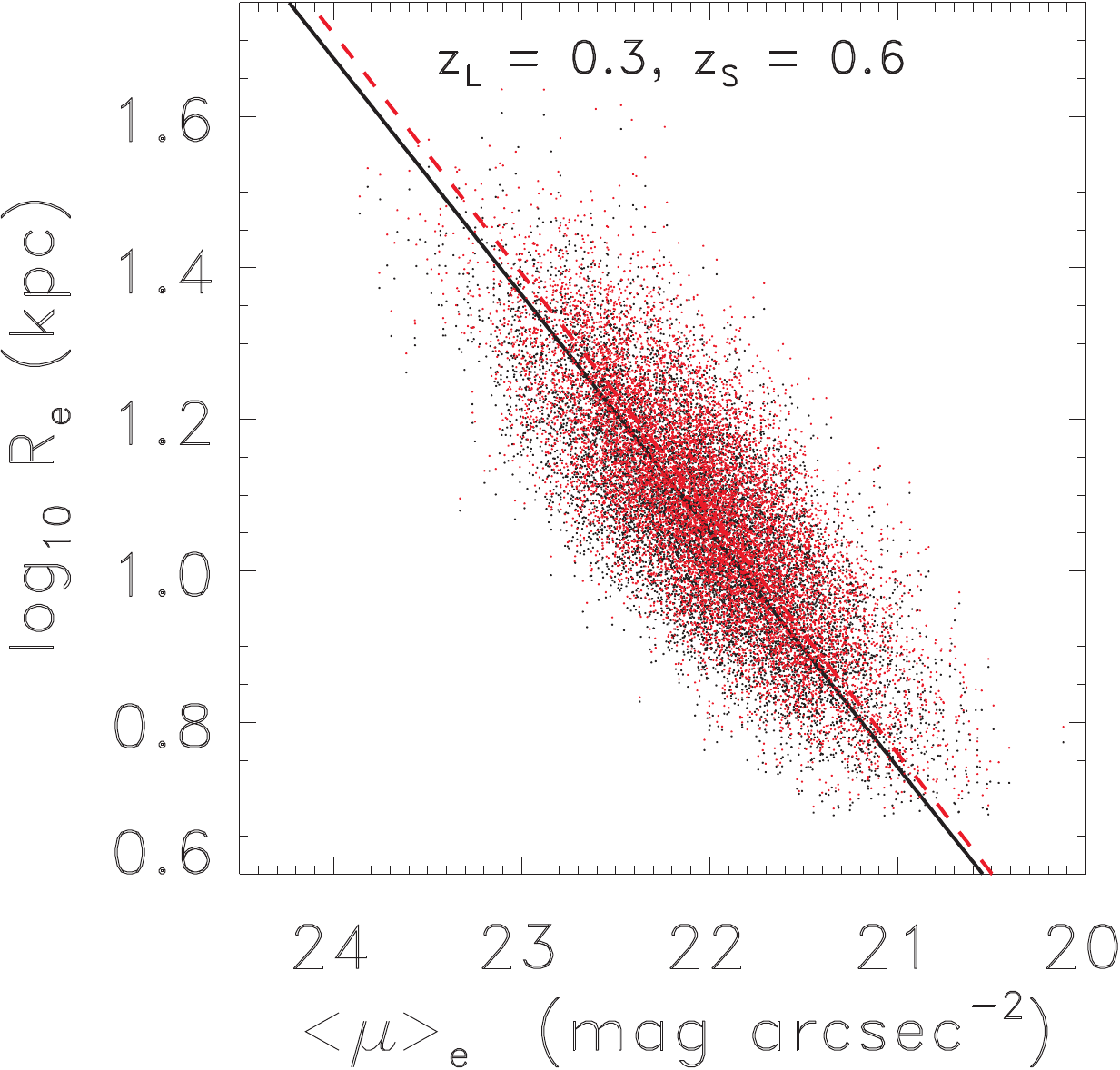}
\end{minipage}
\end{center}
\caption{Kormendy relation of the elliptical galaxies. The
black dashed line shows the Kormendy relation of the unlensed background
elliptical galaxies (black dots). The red dashed line presents the relation of the
lensed elliptical galaxies (red dots). In this figure, the lens galaxies are 
at redshift $z = 0.3$ and the source galaxies at $z=0.6$.
}
 \label{fig:Kormendy_Relation}
\end{figure}

To derive the systematic shifts in the fundamental plane, in principle
we need to measure the velocity dispersion from spectroscopic
observations. This will be very time consuming given the large
number of early-type galaxies available from LSST. However, it is 
easier to measure the \citet{kor77} relation between the surface
brightness and effective radius, which can be 
obtained from imaging data and an approximate photometric redshift. In Fig. \ref{fig:Kormendy_Relation}, we 
show the Kormendy relation, of the lensed and unlensed elliptical
galaxies. On average, an upward offset of about 0.03 in the
Kormendy Relation is seen in Fig. \ref{fig:Fundamental_Plane}. 
As might be expected, this shift is similar to that in the fundamental plane. This offset
is much smaller than the scatter orthogonal to the Kormendy relation (about 0.166) and the scatter in the direction of $\log R_{\rm e}$  (about 0.182). 

\section{Conclusion and Discussion}
\label{sec:conclusion}

In this paper we have investigated the moderate lensing of background
elliptical galaxies by  intervening elliptical galaxies. We find the 
the optical depth for moderate lensing is on the order of $\sim 1\%$.
We have also performed image simulations 
based on the design specifications of LSST and obtained realistic lensed images of the
background elliptical galaxies. The 
distortions of the lensed images have been quantified 
with the IRAF task {\tt ELLIPSE}. We find moderately lensed
galaxies can be potentially differentiated from normal galaxies as outliers
in the coefficients of $a_3/a, b_3/a$ etc. 

We also explore the offsets in the Kormendy relation and fundamental plane of
the lensed elliptical galaxies, and attribute these to the
magnification effect of the effective radii, while the mean surface
brightnesses within the radii are nearly unaffected. The systematic
offsets are  on the order of $0.03$ in $\log R_{\rm e}$.  To observe the shift in the fundamental plane, we need to determine the velocity dispersion spectroscopically while for the Kormendy relation, we only need the source redshift (e.g., from photometric redshift) to obtain the physical effective radius, so the observational demand is somewhat lower. 

For strong lensing, it is possible to model an individual lens to extract information about the lens. For moderate lensing, this will be difficult, and its application will be mostly statistical. Since the scatter in the fundamental plane is small (0.08 in $\log_{10} R_{\rm e}$), a 0.03 shift can in 
principle be deteced with a few thousand galaxies. For the Monte Carlo simulations we discussed in \S\ref{sec:model}, we first fit the unlensed sample with a linear model and then renormalise the error bars so that the total $\chi^2$ per degree of freedom is 
unity. We then fit the same model to the lensed sample. If there were no systematic shifts, then the resulting $\chi^2$ should follow roughly a normal distribution with mean equal to $\chi^2_{\rm mean}=N_{\rm dof}=N-2$ and a dispersion of $\sigma_{\chi}=\sqrt{2 N_{\rm dof}}$, where $N$ is the number of pairs of galaxies. Any significant deviation will indicate the no offset model is unsatisfactory. For our simulated sample, we find the $\chi^2$ for the no offset model is significantly higher than this expectation, and the excess $\chi^2$ above $\chi^2_{\rm mean}$ is on the order of $4.2\sigma_{\chi}$ and $6.6\sigma_{\chi}$ for 1600 and 5000 pairs of galaxies. For the Kormendy relation, the corresponding significance levels are $3.4\sigma_{\rm chi}$ and $6.3\sigma_{\rm chi}$, somewhat lower than that in the fundamental plane, which is not surprising given the larger scatters in this relation.

In practice, there will be several
complications since the lens and source redshifts  
may only be available photometrically. However, if there are no
systematic errors, we can in principle bin the background galaxies,
and stack them to find the systematic offset in the fundamental plane
and Kormendy relation as a function of separation, which in turn provides strong constraints
on the average profiles of galaxies at large radii. This complements
the weak galaxy-galaxy method using shear.

At redshift 0.3, the median Einstein radius is around $0.83$ arcsec (corresponding to about $2.6 h^{-1}$\,kpc), and moderate lensing can probe to $\sim 5\theta_E$, about $13.2 h^{-1}$\,kpc.  This is an interesting radius, since it may be close to the regime where the density slope may be changing from isothermal ($\rho \propto r^{-2}$, \citealt{koo09}) to steeper profiles ($r^{-3}$):  for a galactic-sized halo with virial radius of $r_v=200$ kpc and a concentration parameter of $c=10$, the radius where the density slope changes may be around $r_v/c=20$ kpc.
 
Although the results presented here are promising, there are several points 
which need to be investigated further in the future. First, we do not fully consider the effects of
seeing, as would be necessary in more realistic
simulations. However, some of the quantities that we use, such
as the Fourier components (see Fig. \ref{fig:a3a4b3b4_2}), are already
averaged over a range of radii and so the impact will be somewhat limited.
Second, while many of the 
elliptical galaxies can be described well by the de Vaucouleurs
profile,  others are better described by the more general S\'ersic (1968)
profile. Third, for simplicity we adopt
the singular isothermal lens model. In reality, several other models
can better describe the lens galaxies. These include the singular isothermal ellipsoid model \citep{Keeton98}, 
and  the GNFW model \citep{zha96, chae02} which may be particularly suitable
for studying the transition in the density profiles. In more detailed studies we could parameterise the foreground lens using these models and extract the best-fit parameters with more rigourous statistical methods, such as maximum likelihood or Bayesian techniques. It may be particular interesting to explore what we can learn with photometric redshifts alone for moderate gravitational lensing using Monte Carlo simulations.

\section*{Acknowledgments}
We thank Drs. Cheng Li, Hu Zhan and Zuhui Fan for helpful
discussions. JW and SM acknowledge the Chinese Academy of
Sciences and NSFC (grants 10821061 and 11033003) for financial support.
MCS acknowledges financial support from the Peking University One
Hundred Talent Fund (985) and NSFC grants 11043005 and 11010022
(International Young Scientist).


\begin{thebibliography}{99}
\bibitem[\protect\citeauthoryear{Barkana}{1998}] {bar98} Barkana R. 1998, ApJ, 502, 531
\bibitem[\protect\citeauthoryear{Bernardi et al.}{2003a}] {ber03a} Bernardi et al. 2003a, AJ, 125, 1849
\bibitem[\protect\citeauthoryear{Bernardi et al.}{2003b}] {ber03b} Bernardi et al. 2003b, AJ, 125, 1866
\bibitem[\protect\citeauthoryear{Blanton et al.}{2003}] {bla03} Blanton M. R. et al., 2003, ApJ,
  592, 819 
\bibitem[Browne et al.(2003)]{bro03} Browne I.~W.~A., et 
al.\ 2003, MNRAS, 341, 13 
\bibitem[\protect\citeauthoryear{Bullock et al.}{ 2001}] {bul01} Bullock J. S. et al., 2001,
  MNRAS, 321, 559
  \bibitem[Chae(2002)]{chae02} Chae, K.-H.\ 2002, \apj, 568, 500 
\bibitem[\protect\citeauthoryear{Choi et al.} {2007}] {cho07} Choi Y. Y., Park C., \& Vogeley
  M. S., 2007, ApJ, 658, 884
\bibitem[\protect\citeauthoryear{Davis, Huterer \& Krauss}{2003}] {dav03} Davis A. N., Huterer
  D. \& Krauss L. M., 2003, 344, 1029
  \bibitem[Futamase et al.(1998)]{fut98} Futamase, T., Hattori, 
M., \& Hamana, T.\ 1998, ApJ, 508, L47 
\bibitem[\protect\citeauthoryear{Fukugita, Futamase \& Kasai}{1990}] {Fuk90} Fukugita, M.,
  Futamase, T., \& Kasai, M., 1990, MNRAS, 246, P24 
\bibitem[\protect\citeauthoryear{Gott, Park \& Lee}{1989}] {Gott89} Gott, J. R., Park, M.-G., \& Lee, H.M 1989, ApJ, 338, 1 
\bibitem[\protect\citeauthoryear{Graham et al.}{2005}] {Graham05} Graham, A. W., Driver, S. P., Petrosian, V., Conselice, C. J., \& Goto, T. 2005, ApJ, 130, 1535 
\bibitem[\protect\citeauthoryear{Hao et al.}{2006}] {Hao06} Hao,
  C. N., Mao, S., Deng, Z. G., Xia, X. Y. \& Wu Hong 2006, MNRAS, 370, 1339 
\bibitem[\protect\citeauthoryear{Hogg}{1999}] {hog99} Hogg D. W. 1999, arXiv:9905116
\bibitem[\protect\citeauthoryear{Jedrzejewski}{1987}] {Jedrzejewski87} Jedrzejewski R. I., 1987,
  MNRAS, 226, 747 
\bibitem[\protect\citeauthoryear{Kayser et al.} {1986}] {kay86} Kayser R., Refsdal S., Stabell
  R., 1986, A\&A, 166, 36
\bibitem[\protect\citeauthoryear{Keeton \& Kochanek}{1998}] {Keeton98} Keeton C. R., Kochanek
  C. S., 1998, ApJ, 495, 157 
\bibitem[\protect\citeauthoryear{Komatsu et al.}{2009}] {kom09} Komatsu E. et al., 2009, ApJs,
  180, 330 
  \bibitem[\protect\citeauthoryear{Koopmans}{2009}]{koo09} Koopmans, L.~V.~E., et 
al.\ 2009, ApJ, 703, L51  
\bibitem[\protect\citeauthoryear{Kormendy}{1977}]{kor77} Kormendy J.,
  1977, ApJ, 218, 333 
\bibitem[\protect\citeauthoryear{Li \& Ostriker}{2002}] {li02} Li
  L. X., Ostriker J. P., 2002, ApJ, 566, 652
\bibitem[\protect\citeauthoryear{Loeb \& Peebles}{2003}] {loe03} Loeb A., Peebles P. J. E., 2003, ApJ, 589, 29
\bibitem[\protect\citeauthoryear{McMaster et al.}{2008}] {mcm08}
  McMaster, Biretta, et al. 2008, WFPC2 Instrument Handbook, Version
  10.0 (Baltimore: STScI) 
\bibitem[\protect\citeauthoryear{Navarro, Frenk \& White}{1996}] {nav96} Navarro J. F.,
  Frenk C. S. \& White S. D. M., 1996, ApJ, 462, 563
\bibitem[\protect\citeauthoryear{Oguri et al.}{2002}] {ogu02} Oguri M., Taruya A., Suto Y.,
  Turner E. L., 2002, ApJ, 568, 488
  \bibitem[Oohama et al.(2009)]{ooh09} Oohama, N., Okamura, S., 
Fukugita, M., Yasuda, N., \& Nakamura, O.\ 2009, \apj, 705, 245 
\bibitem[Press et al.(1992)]{press92} Press, W.~H., Teukolsky, 
S.~A., Vetterling, W.~T., 
\& Flannery, B.~P.\ 1992, Cambridge: University Press, |c1992, 2nd ed.,  
\bibitem[\protect\citeauthoryear{Renzini}{2006}] {Renzini06} Renzini A., 2006, ARA\&A, 44, 141
\bibitem[\protect\citeauthoryear{Schechter}{1976}] {sch76} Schechter P., 1976, ApJ, 203, 297
\bibitem[\protect\citeauthoryear{Schneider et al.}{1992}] {sch92} Schneider P., Ehlers J., Falco E. E., 1992, Gravitational Lenses (Springer Verlag, Berlin)
\bibitem[\protect\citeauthoryear{Scodeggio et al.}{1998}] {sco98}
  Scodeggio M., Gavazzi G., Belsole E., Pierini D., Boselli A., 1998,
  MNRAS, 301, 1001 
\bibitem[\protect\citeauthoryear{Sheth et al.}{2003}] {she03} Sheth et al., 2003, ApJ, 594, 225
\bibitem[\protect\citeauthoryear{Sonnenfeld et al.}{2011}]{son11} Sonnenfeld, A., 
Bertin, G., \& Lombardi, M.\ 2011, arXiv:1106.1442 
\bibitem[\protect\citeauthoryear{Turner}{1980}] {tur80} Turner E. L., 1980, ApJ, 242, L135
\bibitem[\protect\citeauthoryear{Turner et al.}{1984}] {tur84} Turner E. L., Ostriker J. P., \&
  Gott J. R., III. 1984, ApJ, 284, 1
 \bibitem[\protect\citeauthoryear{Schneider et al.}{2006}]{skw06} Schneider, P., Kochanek, C.~S., \& Wambsganss, J.\ 2006, Gravitational Lensing: Strong, Weak and Micro: , Saas-Fee Advanced Courses, Volume 33.~ISBN
978-3-540-30309-1.~Springer-Verlag (Berlin) 
\bibitem[\protect\citeauthoryear{Seitz et al.}{1998}]{sei98} 
Seitz S., Saglia R.~P., Bender R., Hopp U., Belloni P., Ziegler B., 1998, 
MNRAS, 298, 945 
 \bibitem[\protect\citeauthoryear{Sersic}{1968}]{ser68}  Se«rsic,
   J.-L. 1968, Atlas de Galaxias Australes (Co« rdoba: Obs. Astron.) 
\bibitem[\protect\citeauthoryear{Williams 
\& Lewis}{1998}]{wil98} Williams L.~L.~R., Lewis G.~F., 1998, MNRAS, 294, 299 
\bibitem[\protect\citeauthoryear{Wyithe et al.}{2010}] {wyi10} Wyithe J. S. B., Oh S. P., Pindor
  B., 2010, arXiv:1004.2081v1
\bibitem[\protect\citeauthoryear{Wyithe et al.}{2003}] {wyi03} Wyithe J. S. B., Winn J. N.,
  Rusin D., 2003, ApJ, 583, 58
\bibitem[\protect\citeauthoryear{Wyithe et al.}{2001}] {whi01} Wyithe J. S. B., Turner E. L.,  Spergel D. N., 2001, ApJ, 555, 504
\bibitem[Yee et al.(1996)]{yee96} Yee, H.~K.~C., Ellingson, 
E., Bechtold, J., Carlberg, R.~G., \& Cuillandre, J.-C.\ 1996, \aj, 111, 1783 
\bibitem[\protect\citeauthoryear{Zhao}{1996}] {zha96} Zhao H. S., 1996, MNRAS, 278, 488
\end{thebibliography}
\end{document}